\documentclass{article}  
\usepackage{bigsky2009}
\usepackage{graphicx}
\usepackage{amssymb}
\frompage{000} \topage{000}                                              
\def\rightmark{Universal Flow in the Early Stages at RHIC}
\def\leftmark{S. Pratt}
 
\title{Universal Flow in the Early Stages at RHIC} 
\authors{
{Scott Pratt %
}\\[2.812mm]
{\normalsize
\hspace*{-8pt}Department of Physics \&Astronomy,Michigan State University,\\
East Lansing, MI 48824 ~~USA\\[0.2ex] 
}}
 
\abstract{Pre-thermal flow plays an important role in the final state evolution of heavy ion collisions at RHIC. We show that flow has universal features for a wide range of models. This significantly reduces the uncertainty in initializing hydrodynamic models at RHIC.}


\begin{document}
 
\maketitle
\setcounter{page}{1}

\section{Introduction and Basic Theory}\label{intro}

Flow plays a pivotal role in driving soft observables at RHIC, including spectra, elliptic anisotropies, and two-particle correlations. For a thermalized system, flow is driven by the equation of state and viscosity, and thus understanding collective flow is synonymous with our ability to discern bulk properties of the novel matter created in heavy ions at RHIC. However, significant amounts of flow are created in the final stages of the reaction, $\tau\gtrsim 7$ fm/$c$, or the pre-thermalized stage, $\tau\lesssim 1$ fm/$c$. For the latter stages, the degrees of freedom are hadronic, and the interactions become largely binary, which allows one to model the latter stage rather confidently with hadronic cascades. However, there is significant debate regarding the microscopic form of matter during the pre-thermalized stage. Early flow has been shown to be pivotal in reproducing femtoscopic two-particle correlation data from hydrodynamic models \cite{Gyulassy:2007zz,Li:2007yd,Pratt:2008sz,Broniowski:2008vp}. The matter might be dominated by partonic degress of freedom, which may be far from kinetically thermalized, to classical color fields, which carry no entropy. The stress-energy tensors, $T_{\alpha\beta}$, for these different pictures, are very different. The transverse pressure $T_{xx}$, could be $T_{00}/3$ for thermalized massless partons, $T_{00}/2$ for non-interacting partons, or approach $T_{00}$ for longitudinal color fields. The longitudinal pressure, $T_{zz}$, varies correspondingly, thus leading to very different evolutions of the energy density during the initial fm/$c$. In fact, for a given energy density $T_{00}$, at $\tau=0.1$ fm/$c$, the energy density can easily differ by a factor of 2 by $\tau=1$ fm/$c$, depending on the state of matter during the pre-thermalized stage. 

In contrast to the evolution of the energy density, the collective transverse flow, as defined by 
\begin{equation}
{\rm Flow}\equiv\frac{T_{0i}}{T_{00}},
\end{equation}
varies very little for the different pictures \cite{Vredevoogd:2008id}. Since the transverse pressures differ by a factor of 3 in the pictures mentioned above, this statement seems counter-intuitive. To understand the statement, we first note that all the pictures above share the property that the stress-energy tensor is traceless. During the first fm/$c$, energy densities typically surpass 10 GeV/fm$^3$, which is sufficiently far above $T_c$ to warrant this statement, at least at the 10\% level. Since $T_{xx}+T_{yy}+T_{zz}=\epsilon$ at early times, and since $T_{xx}=T_{yy}$, one can describe the initial stress-energy tensor by two numbers, $\epsilon$ and $\kappa$, with 
\begin{equation}
T_{xx}=T_{yy}=\kappa T_{00},~~~T_{zz}=(1-2\kappa)T_{00}.
\end{equation}
Here, $\kappa$ represents the transverse stiffness, with $\kappa=1/3$ for ideal hydrodynamics, $\kappa=1/2$ for free-streaming particles, and $\kappa=1$ for purely longitudinal electric fields, or purely longitudinal magnetic fields. Energy and momentum conservation give:
\begin{eqnarray}
\partial_\tau T_{00}&=&\partial_zT_{0z}/\tau,\\
\nonumber
\partial_\tau T_{0x}&=&\partial_xT_{xx}+\partial_zT_{xz}.
\end{eqnarray}
At $z=0$, $T_{0z}$ and $T_{xz}$ vanish, but for small $z$ longitudinal collective flow ($v_z=z/t$),
\begin{equation}
\label{eq:smallz}
T_{0z}=(T_{00}+T_{zz})z/\tau,~~~T_{xz}=T_{0x}z/\tau,
\end{equation}
which leads to the small time behavior for $T_{0x}$ and $T_{00}$,
\begin{eqnarray}
\partial_\tau T_{0x}&=&-\kappa\partial_xT_{00}-\frac{T_{0x}}{\tau},\\
\nonumber
\partial_\tau T_{00}(\tau)&=&-\frac{1}{\tau}(2-2\kappa)T_{00}.
\end{eqnarray}
Combining these, one finds that the flow grows as:
\begin{eqnarray}
\label{eq:flow1}
\partial_\tau\frac{T_{0x}}{T_{00}}&=&\frac{\partial_\tau T_{0x}}{T_{00}}
-\frac{T_{0x}}{T_{00}}\frac{\partial_\tau T_{00}}{T_{00}}\\
\nonumber
&=&-\frac{\kappa\partial_x T_{00}+T_{0x}/\tau}{T_{00}}
+\frac{(2-2\kappa)}{\tau}\frac{T_{0x}}{T_{00}}.
\end{eqnarray}
Defining $\alpha$ to describe the linear growth of the flow at small times,
\begin{equation}
\alpha\equiv \partial_\tau\left(\frac{T_{0x}}{T_{00}}\right),~
~{\rm or}~~\frac{T_{0x}}{T_{00}}=\alpha_x\tau,
\end{equation}
one can insert these expressions into Eq. (\ref{eq:flow1}) and solve for $\alpha$,
\begin{equation}
\label{eq:universal}
\alpha_x=-\frac{\partial_xT_{00}}{2T_{00}}.
\end{equation}
This last expression is referred to as ``universal flow''. The universality comes from the fact that $\alpha_x$ is independent of $\kappa$, meaning that the initial growth of the flow does not depend on the transverse stiffness, and will be identical for the very different pictures mentioned above. However, universality is contingent on three criteria:
\begin{enumerate}
\item The stress-energy tensor is traceless. This is true for non-interacting gauge fields, or non-interacting massless partons, and does not require thermalization. For the energy densities in the first fm/$c$, mostly above 10 GeV/fm$^3$, this is true to the order of 10\%. For lower energy densities in the critical region, this is a poor assumption.
\item The longitudinal motion must respect boost invariance. This should be a good assumption for RHIC energies and an excellent assumption at LHC energies. The most important part of the assumption is that the collective velocity is $u_z=z/\tau$, which was applied in Eq. (\ref{eq:smallz}).
\item The transverse stiffness $\kappa$ must be independent of $x$ and $y$, though it can depend on $\tau$. Otherwise, $\partial_x T_{xx}$ could not be identified as $\kappa\partial_xT_{00}$.
\end{enumerate}

\section{Models}

If universality holds for sufficient time, it can be used to initialize the flow in hydrodynamic models. For the Navier-Stokes limit of viscous hydrodynamics, the stress-energy tensor is determined by four pieces of information: the energy density $\epsilon$, and collective velocities $u_i$, or equivalently $T_{00}$ and $T_{0i}$. However, more generally, the stress-energy tensor has 10 free parameters, as the spatial components $T_{ij}$ might differ from their Navier-Stokes values if not given sufficient time to relax to their correct values. This relaxation is included in Israel-Stewart approaches and is characterized by a relaxation time $\tau_{IS}$ \cite{israelstewart,muronga,baierromatschke,heinzchaudhuri,romatschke,songheinz}. This relaxation time is determined by the viscosity and the pressure fluctuation \cite{Pratt:2007gj,jou},
\begin{eqnarray}
\eta=\frac{\tau_{IS}}{T}\int d^3r\langle T_{xy}(0)T_{xy}(r)\rangle.
\end{eqnarray}
The pressure fluctuation can be estimated by considering a quasi-particle picture of partons, or in principle from Lattice Gauge theory without the need to invoke analytic continuation. Assuming that the viscosity is within a factor of 3-4 or less of the KSS limit, one would expect $\tau_{IS}\approx 0.5$ fm/$c$. This suggests that the Navier-Stokes picture should become valid at $\tau\lesssim 1$ fm/$c$. Thus, if universal flow can last all the way to $\tau=1$ fm/$c$, one could confidently initialize the collective flow for the hydrodynamic calculation independent of the model, beginning at $\tau=1$ fm/$c$.

\begin{figure}
\centerline{\includegraphics[width=0.6\textwidth]{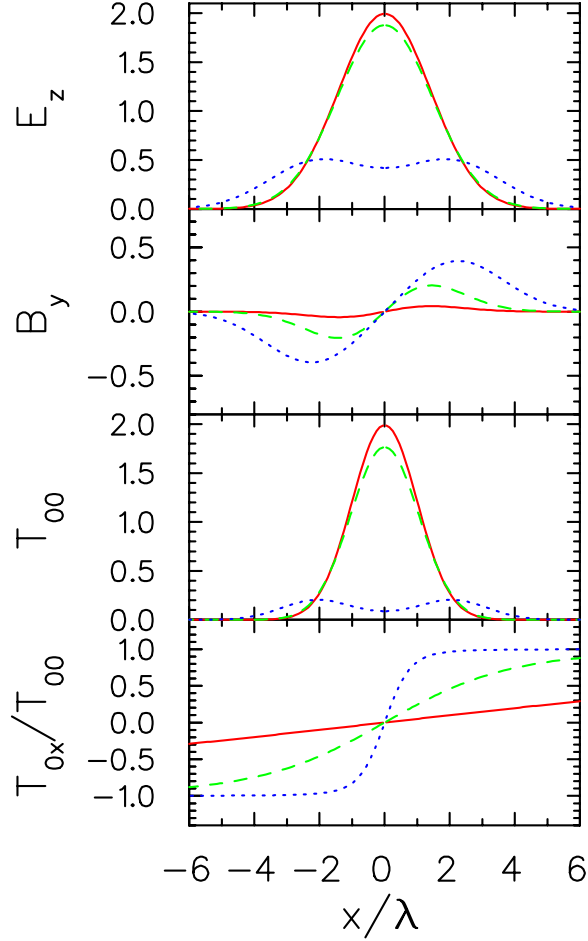}}
\caption{\label{fig:fluxtube}
The evolution of the electromagnetic fields and flux tubes from two oppositely Gaussian charge densities receding at $\pm c$ for three proper times: $\tau=0.1\lambda$ (red lines), $\tau=0.5\lambda$ (green dashed lines) and $\tau=2.5\lambda$ (blue dotted lines), where $\lambda$ characterizes the transverse Gaussian size of the radially symmetric charge densities. Each quantity is plotted as a function of the position along the $x$ axis. The electric field remains longitudinal and the magnetic field curve around the $z$ axis. The energy density $T_{00}$ spreads out over time, developing into a pulse for $\tau>>\lambda$. The flow $T_{0x}/T_{00}$ grows linearly before its magnitude saturates at unity.}
\end{figure}
In order to test this possibility, we compare three very different models. The first two models will be based on non-interacting fields, generated by charged capacitor plates flying apart at $\pm c$, with opposite charge densities $\pm \rho(x,y)$. From these currents, one can generate the electromagnetic fields,
\begin{eqnarray}
E_z(x,y,\tau)&=&\int dx'dy'~\rho(x',y')\delta((x-x')^2+(y-y')^2-\tau^2),\\
\nonumber
{\bf B}(x,y,\tau)&=&\int dx'dy'\rho(x',y')\delta((x-x')^2+(y-y')^2-\tau^2)\frac{({\bf r}-{\bf r}')\times\hat{z}}{|{\bf r}-{\bf r}'|}.
\end{eqnarray}
For a single Gaussian charge density, $\rho(r)=e^{-r^2/4\lambda^2}$, the resulting evolution is displayed in Fig. \ref{fig:fluxtube}. Here, $\lambda$ is the color coherence length, or inverse saturation scale. For $\tau<<\lambda$, a longitudinal electric field dominates the stress-energy tensor, and $T_{xx}=T_{yy}=T_{00}$, while $T_{zz}=-T_{00}$. On a time scale of $\lambda$, the magnetic field builds up, and along with it a momentum density, $T_{0x}=E_zBy$. For large times, the field develops into a pulse of radius $\tau$ and thickness $\lambda$, and the effective $\kappa$ approaches 1/2. Greatly more sophisticated calculations of interacting color fields reveal similar evolutions of the transverse stiffness \cite{Krasnitz:2002mn}.

Each of the three models will have an initial energy density with the same transverse profile,
\begin{equation}
T_{00}(x,y,\tau\rightarrow 0)\sim e^{-(x^2+y^2)/2R^2},
\end{equation}
with $R=3$ fm. The first model considers a single coherent charge across the entire transverse area. This is clearly unphysical for a color charge, but is chosen to represent an extreme limit as it maintains $\kappa\approx 1$ for the longest time. The second model is to consider an ensemble of point-like charges. For this case $\kappa$ is 1/2 for the entire evolution. As a third model, we consider ideal thermodynamics of a massless gas, and $\kappa=1/3$.

\begin{figure}
\begin{minipage}[t]{0.5\textwidth}
\centerline{\includegraphics[width=0.95\textwidth]{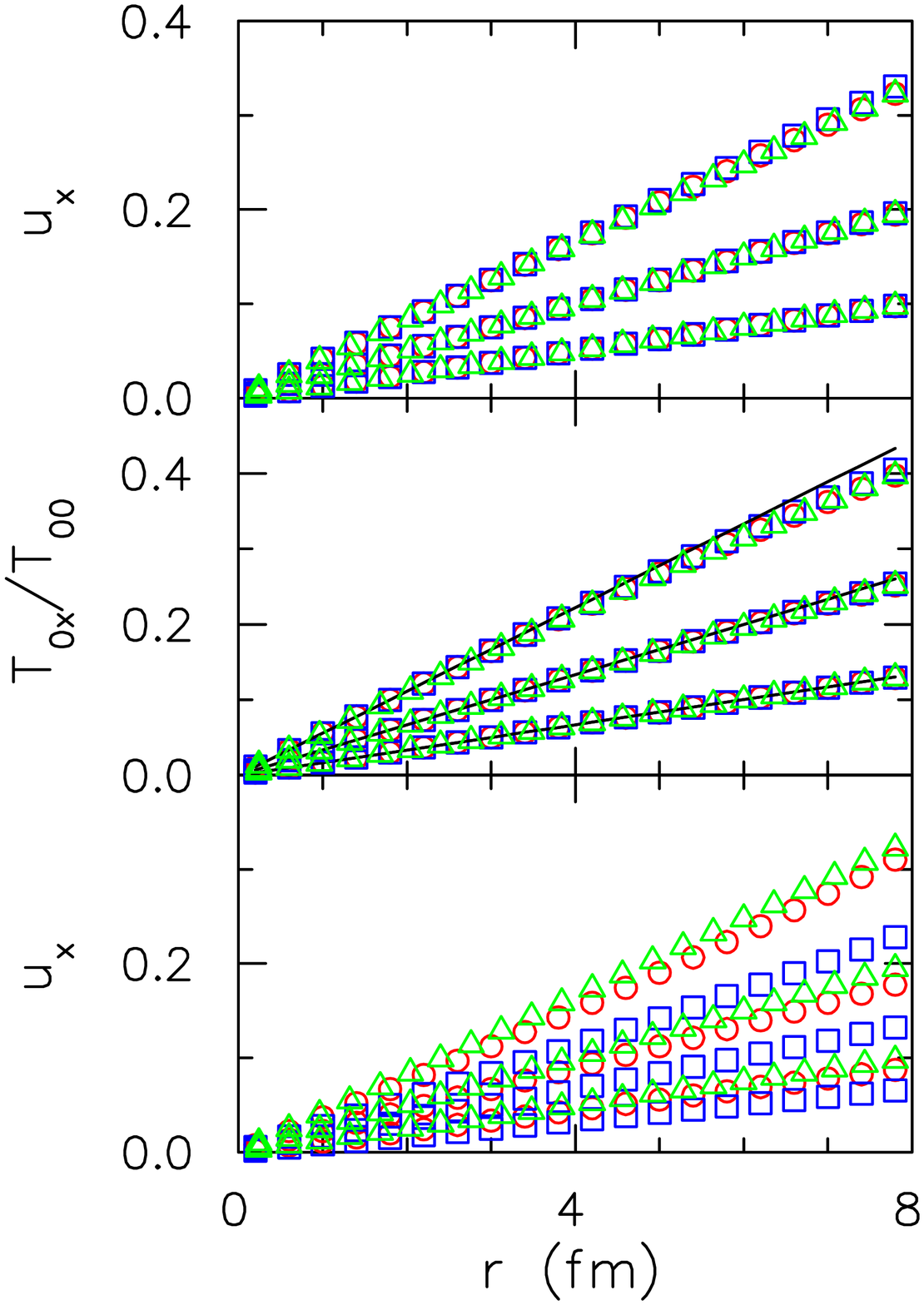}}
\caption{\label{fig:universal}
Results for the three models at three different times, 0.3, 0.6 and 1.0 fm/$c$ as a function of the position along the $x$ axis. The three models described in the text are: Ideal hydrodynamics (green triangles), and the coherent (blue squares) and incoherent (red circles) limits of classical fields.\newline
Lower Panel:
Ideal hydrodynamics has the greatest transverse radial collective velocity $u_x$ even though it has the smallest transverse pressure, $T_{xx}$, of all three models. 
\newline
Middle Panel: The flow, $T_{0x}/T_{00}$, is nearly universal for all three models.\newline
Upper Panel: The collective velocity assuming that the matter suddenly behaves as if it were ideal hydrodynamics at the prescribed time. Since this is determined by fitting $T_{0x}/T_{00}$, it is also nearly universal. If the fit had been to a Navier Stokes form of viscous hydrodynamics, the forms would also have converged, but to different lines.
}
\end{minipage}
\begin{minipage}[t]{0.5\textwidth}
\centerline{\includegraphics[width=0.95\textwidth]{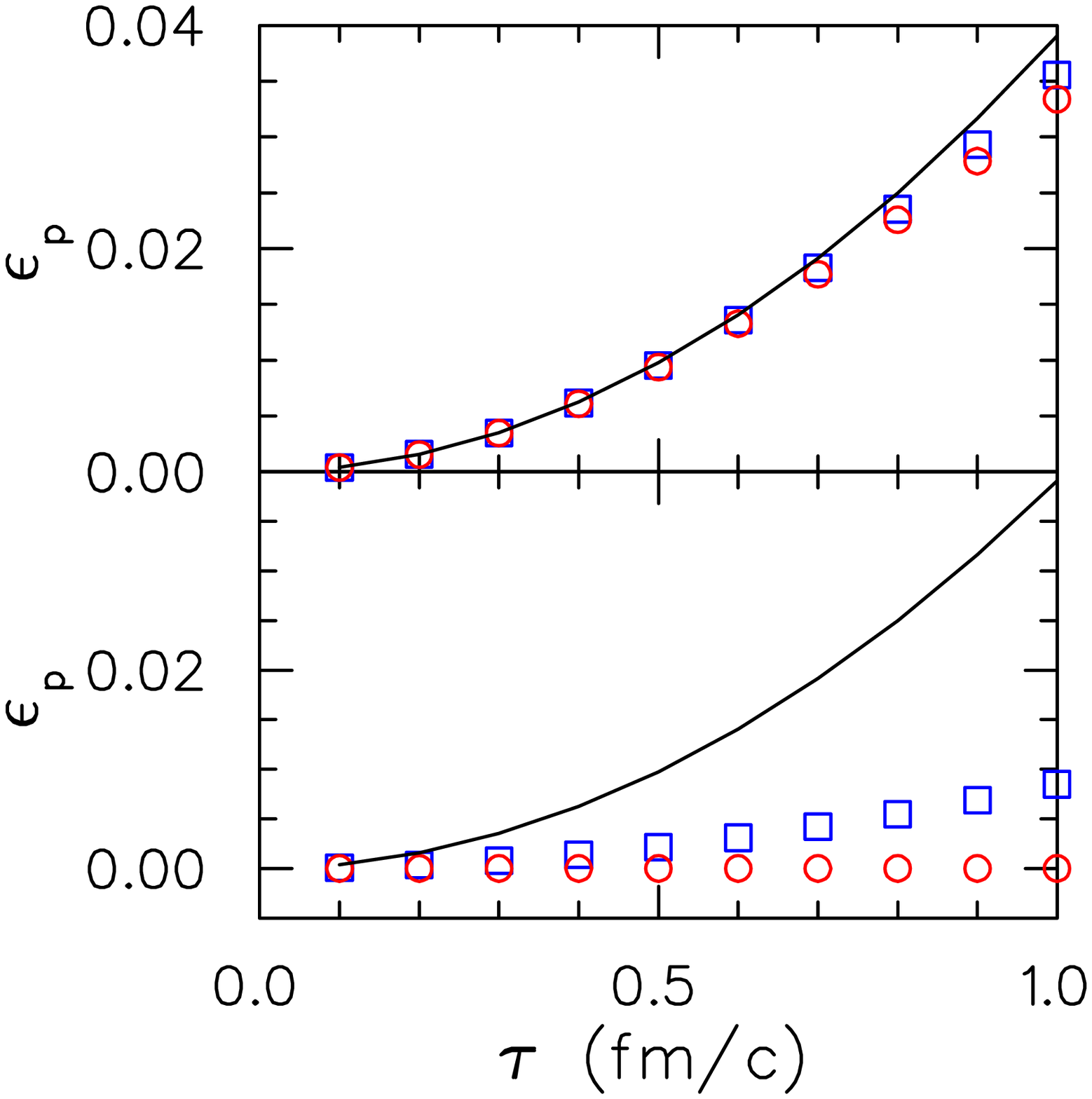}}
\caption{\label{fig:epsilonp}
Elliptic anisotropies, defined by Eq. (\ref{eq:epsilonpdef}), as a function of the proper time for a source with initial Gaussian energy density characterized by $R_x=2$ and $R_y=3$ fm.
Lower Panel:  $\epsilon_p$ for  coherent (blue squares) and incoherent (red circles) limits of non-interacting fields. The incoherent case yields $\epsilon_p=0$, exactly as one would obtain with non-interacting particles. The solid line shows the lowest-order (in $\tau$) quadratic contribution for ideal hydrodynamics.\newline
Upper Panel: Assuming that the matter suddenly behaves as if it were ideal hydrodynamics at time $\tau$, $\epsilon'_p$ represents the anisotropy of the altered stress-energy tensor, after fitting to $T_{00}$, $T_{0x}$ and $T_{0y}$. The result is close to the quadratic form approximating the behavior of ideal hydrodynamics. If one had matched to a Navier-Stokes form, rather than an ideal form, the three solutions would have also converged, but to a different line.
}
\end{minipage}
\end{figure}

Figure \ref{fig:universal} shows the flow $T_{0x}/T_{00}$ along the $x$ axis for three different times for each of the three models, and for the analytic answer of Eq. (\ref{eq:universal}). Indeed, all four models are very similar, even for the larger time of 1.0 fm/$c$. Also shown are the collective velocities $u_x$, which are defined by the four-velocity required to make $T_{0x}$ vanish in the observer's frame. The velocities differ substantially between the models. If one were to thermalize the systems suddenly at the chosen times, one could find the new velocities required to maintain the energy and momentum densities while matching the hydrodynamic form of the stress-energy tensor. The velocities then suddenly change, and once again match each other, as illustrated in the upper panel. This emphasizes the impact of making the suddenness in the laboratory frame, rather than in the frame of the fluid. Otherwise, $\epsilon$ and $u_i$ would be preserved rather than $T_{00}$ and $T_{0i}$. Making the change locally simultaneous according to a co-moving observer would insinuate an inside-out transition, with the inside thermalizing before the outside. The sensitivity to reference frame for changing $\kappa$ is related to the third assumption for universality, i.e., $\kappa$ depends only on $\tau$ and not on $x$ or $y$.

The success of matching the flow for all three very different models all the way to $\tau=1$ fm/$c$, suggests that
one can rather confidently use Eq. (\ref{eq:universal}) to initialize the initial flow for a hydrodynamic or viscous hydrodynamic code in a way that is insensitive to the anisotropy of the stress-energy tensor at earlier times. For a viscous code, the initial collective velocity $u_i$ would depend on the viscosity one chooses to apply at one fm/$c$, i.e., the values $u_i'$ shown in the upper panel of Fig. \ref{fig:universal} would change, but they would all change to the same curve.

Elliptic flow can also build up in the pre-thermal stage. Figure \ref{fig:epsilonp} shows the evolution of 
\begin{equation}
\label{eq:epsilonpdef}
\epsilon_p\equiv\frac{\int dxdy~(T_{xx}-T_{yy})}{\int dxdy~(T_{xx}+T_{yy})},
\end{equation}
which can be identified with the elliptic flow once the system has become free-streaming particles. Since $\epsilon_p$ depends on $T_{xx}$ and $T_{yy}$, it varies dramatically between the three models. However, if the systems were to suddenly thermalize, and one were to match $T_{00}$, $T_{0x}$ and $T_{0y}$ to the hydrodynamic form, $\epsilon_p$ changes suddenly, and the models jump to approximately the same value. For the case of incoherent color fields, which is identical to what one would obtain for free-streaming massless partons, $\epsilon_p$ is zero before thermalization. This comes from the fact that the contribution to the anisotropy from the flow velocity is canceled by the opposite anisotropy in $T_{ij}$ measured by a co-mover. If the system suddenly thermalizes, this negative asymmetry disappears and leaves a postive value for $\epsilon_p$ This demonstrates that the conditions for elliptic flow can be generated even while $\epsilon_p$ is zero.

Universality has important implications for the modeling of heavy ion collisions at high energy. Running a viscous hydrodynamic code requires choosing an initial energy profile and an initial flow. The findings here show that the uncertainty of the flows are largely ameliorated by universality, but one should not forget that the initial energy density is certainly uncertain, and varies drastically for the pictures discussed here. For the coherent limit, the energy density stays roughly constant for $\tau<<R$, while it fall as $1/\tau$ for the incoherent limit, and as $1/\tau^{4/3}$ for the hydrodynamic limit. Also, the compactness and anisotropies are uncertain at the 10\% level, which translate into 10\% alterations of final state elliptic flow \cite{Drescher:2007cd} and correlations \cite{Broniowski:2008vp,Pratt:2008sz} and spectra. The universality studies shown here allow one to state the flow at 1 fm/$c$ with $\approx 10\%$ confidence, and given that the of the order of 10\% of the flow is built up during this time, one would only expect the uncertainties in the initial flow field to translate to $\approx$ 1\% uncertainty of the final flow.

\section*{Acknowledgments}
Support was provided by the U.S. Department of Energy, Grant No. DE-FG02-03ER41259.

\vfill\eject

\begin{thebibliography}{9}

\bibitem{Gyulassy:2007zz}
  M.~Gyulassy, Yu.~M.~Sinyukov, I.~Karpenko and A.~V.~Nazarenko,
  Braz.\ J.\ Phys.\  {\bf 37}, 1031 (2007).
  
\bibitem{Li:2007yd}
  Q.~Li, M.~Bleicher and H.~Stocker,
  Phys.\ Lett.\  B {\bf 659}, 525 (2008)
  [arXiv:0709.1409 [nucl-th]].

\bibitem{Pratt:2008sz} S. Pratt and J. Vredevoogd, to appear in Phys. Rev. C, arXiv.org, nucl-th 0809.0516.

\bibitem{Broniowski:2008vp} W. Broniowski, M. Chojnacki, W. Florkowski and A. Kisiel, Phys. Rev. Lett. {\bf 101}, 022301 (2008).
  
\bibitem{Vredevoogd:2008id} J. Vredevoogd and S Pratt, arXiv.org, nucl-th 0810.4325 (2008).

\bibitem{israelstewart} W. Israel, Ann. Phys. {\bf 100}, 310 (1976); W. Israel and J.M. Sewart, Ann. Phys. {\bf 118}, 341 (1979).

\bibitem{muronga} A.~Muronga,
  arXiv:0710.3280 [nucl-th] (2007);
  A.~Muronga,
  arXiv:0710.3277 [nucl-th] (2007).
  
\bibitem{baierromatschke} R.~Baier, P.~Romatschke and U.~A.~Wiedemann, Phys.\ Rev.\  C {\bf 73}, 064903 (2006).

\bibitem{heinzchaudhuri}
  U.~W.~Heinz, H.~Song and A.~K.~Chaudhuri,
  Phys.\ Rev.\  C {\bf 73}, 034904 (2006).

\bibitem{romatschke} P.~Romatschke and U.~Romatschke,
  arXiv:0706.1522 [nucl-th] (2007); 
  P.~Romatschke,
  Eur.\ Phys.\ J.\  C {\bf 52}, 203 (2007).

\bibitem{songheinz} H.~Song and U.~W.~Heinz,
  arXiv:0712.3715 [nucl-th] (2007);
  Phys.\ Lett.\  B {\bf 658}, 279 (2008)
  [arXiv:0709.0742 [nucl-th]].
  
\bibitem{Pratt:2007gj}
  S.~Pratt,
  Phys.\ Rev.\  C {\bf 77}, 024910 (2008)
  [arXiv:0711.3911 [nucl-th]].

\bibitem{jou} D. Jou, J. Casas-V\'azquez and G. Lebon, Rep. Prog. Phys. {\bf 51}, 1105 (1988).
  
\bibitem{Krasnitz:2002mn}
  A.~Krasnitz, Y.~Nara and R.~Venugopalan,
  Nucl.\ Phys.\  A {\bf 717}, 268 (2003).
  

\bibitem{Drescher:2007cd}
  H.~J.~Drescher, A.~Dumitru, C.~Gombeaud and J.~Y.~Ollitrault,
  Phys.\ Rev.\  C {\bf 76}, 024905 (2007)
  [arXiv:0704.3553 [nucl-th]].
  


\end{thebibliography}
\end{document}